\documentclass[aps,prc,twocolumn,groupedaddress]{revtex4}

\usepackage{amssymb}
\usepackage{epsfig}

\begin{document}


\title{Three-body resonances $\Lambda nn$ and $\Lambda\Lambda n$}


\author{V. B. Belyaev$^{1,2}$}
\affiliation{$^1$Laboratory of Theoretical Physics, JINR, Dubna,
  141980, Russia}
\author{S. A. Rakityansky$^{1,2}$}
\affiliation{$^2$Department of Physics, University of South Africa, P.O. Box 392,
Pretoria 0003, South Africa}
\author{W. Sandhas$^3$}
\affiliation{$^3$Physikalisches Institut, Universitat Bonn, D-53115
  Bonn, Germany}


\date{\today}

\begin{abstract}
\noindent
Possible bound and resonant states of the hypernuclear systems
$\Lambda nn$ and $\Lambda\Lambda n$ are sought as zeros of the
corresponding three-body Jost functions calculated within the
framework of the hyperspherical approach with local two-body S-wave
potentials describing the $nn$, $\Lambda n$, and $\Lambda\Lambda$
interactions. Very wide near-threshold resonances are found for both
three-body systems. The positions of these resonances turned out to be
sensitive to the choice of the $\Lambda n$-potential. Bound $\Lambda
nn$ and $\Lambda\Lambda n$ states only appear if the two-body
potentials are multiplied by a factor of $\sim 1.5$.\\[3mm]
\begin{tabular}{rcl}
PACS &:& 13.75.Ev, 21.80.+a, 21.45.+v, 25.70.Ef\\
Keywords &:& Lambda-nucleon potential, hypernuclei, three-body resonance,
hyperspherical harmonics,\\ && Jost function
\end{tabular}
\end{abstract}


\maketitle


\section{Introduction}
The $\Lambda$-hyperon belongs to a wide class of particles that are
not in abundance in this world and therefore are not freely available
for scattering experiments. The properties of their interaction with
other particles are studied indirectly. For example, the most
important and established way of studying the $\Lambda N$ interaction
consists in measuring and calculating the spectral properties of the
so called $\Lambda$-hypernuclei (see, for example,
Refs. \cite{gal,nogga} and references therein), which are bound states
of $\Lambda$-particles inside atomic nuclei.  The most convenient for
this purpose are very light nuclei with $A\lesssim10$. Firstly,
because such simple systems have simple spectra with only few well
separated levels, and secondly, because they allow a reliable
theoretical modelling based on rigorous few-body methods.

The hyperon-nucleon attraction is insufficient to bind a $\Lambda N$
pair. The simplest hypernucleus is therefore the hypertriton
${}^3_\Lambda$H, i.e. a bound $\Lambda pn$ complex. Its binding
energy is very small (the $\Lambda$ particle is separated at
$\sim0.15$\,MeV) \cite{miyagawa, nemura1, garcilazo}. So, it looks
like a deuteron core surrounded by a $\Lambda$-halo \cite{miyagawa,
garcilazo}.

Similarly to traditional (non-strange) nuclear physics, where the
deuteron is the first testing ground for any $NN$ potential, the
system $\Lambda NN$ is used to constrain new models of the
hyperon-nucleon interaction. This system was recently analyzed in Refs.
\cite{garcilazo, garcilazo_1} using rigorous three-body equations with the
potentials constructed within the constituent quark model. The authors
of Ref. \cite{garcilazo} gave another proof that the coupling between
the $\Lambda NN$ and $\Sigma NN$ channels is very important for the
hypertriton binding and showed that ${}^3_\Lambda$H is the only bound
state of the $\Lambda NN$ system. Their comprehensive analysis lacks
only one thing: they did not consider possible three-body
resonances. Meanwhile their results give a strong indication that such
resonances may exist and be located not far from the threshold
energy. Indeed, they found that the channel $\Lambda nn$ is attractive
but not sufficient to produce a bound state, and the curve for its
Fredholm determinant turnes towards zero near the threshold energy
(see Fig. 4 of Ref. \cite{garcilazo}).  In our present paper, we
partly fill in the gap by considering the $\Lambda nn$ resonance
state.

The $\Lambda N$ and $\Lambda\Lambda$ potentials are usually
constructed in such a way that the calculations with these potentials
reproduce experimentally known bound states of the hypernuclei.
Unfortunately, it is very difficult to do scattering experiments with
the $\Lambda$-particles because of their short lifetime
($\sim10^{-10}$\,sec) and extremely low intensity of the beams that
can be obtained.

It is well known that even when scattering data are available in full,
it is impossible to construct an interaction potential in a unique
way. One can always obtain different but phase-equivalent potentials
(see, for example, Ref. \cite{belyaevbook}). In this respect the
$\Lambda N$-case is beyond any hope since only few experimental points
for the $\Lambda p$ scattering are available\cite{alexander,
ansari}. During the decades of studying the hypernuclei many features
of the $\Lambda N$-interaction have been revealed. However the
comparison of the theoretical and experimental spectra remains
inconclusive. Different potentials lead to almost the same spectra of
the hypernuclei. We therefore need an additional tool for testing the
potentials.

In principle, such a tool could be based on studying the
$\Lambda$-nucleus resonances, if they do exist
\cite{afnan,kahana}. Indeed, while the scattering and bound states
mostly reflect the on-shell properties of the interaction, the
resonances strongly depend on its off-shell characteristics, which may
be different for phase-equivalent potentials.

Our present work is an attempt to attract the attention of both
theoreticians and experimentalists to the low-energy resonances in the
$\Lambda$-nuclear systems. As an example, we consider the three-body
systems $\Lambda nn$ and $\Lambda\Lambda n$ in the minimal
approximation, $[L]=[L_{\rm min}]$, of the hyperspherical harmonics
approach. By locating the $S$-matrix poles on the second (unphysical)
sheet of the complex energy surface, we show that these systems have
near-threshold resonant states. The position of the poles turnes out
to be strongly dependent on the choice of the $\Lambda N$-potential.
This fact supports the idea that the studying of the
$\Lambda$-nucleus resonances could be very important for finding an
adequate $\Lambda N$-potential.

The demands for adequate hyperon-nucleon ($YN$) and hyperon-hyperon
($YY$) potentials come not only from nuclear physics itself, but also
from astrophysics. The studies of the neutron stars show that these
very dense and compact objects are in fact ``giant hypernuclei'' (see,
for example, Ref. \cite{schaffner} and references therein). The
$\Lambda$-particles appear inside neutron stars when the density
becomes approximately two times higher than the ordinary nuclear
density. The equation of state, describing a neutron star, involves
all the inter-particle potentials and therefore its solutions depend
on their properties. In particular, the strength of the short-range
repulsion in the pairs $\Lambda N$ and $\Lambda\Lambda$ is crucial for
determining the maximum mass and size of a neutron star. The repulsive
nature of the $\Lambda nn$ three-body force (if it is indeed
repulsive) would lead to additional stability of neutron
stars. Moreover, the two-body $YY$ interactions regulate the cooling
behaviour of massive neutron stars \cite{schaffner}.

So, the studies of hypernuclear systems are not only important for
reaching a better understanding of the physics of strange particles,
but may also have an important impact on some other branches of
science.  This is why the research in this field is carried on by many
theoretical groups and experimental laboratories.

\section{Three-body Jost function}

There are several different ways of locating quantum resonances. The
most adequate are the methods based on the rigorous definition of
resonances as the $S$-matrix poles at complex energies. This
definition is universal and applicable to the systems involving
more than just two colliding particles. Of course, the problem of
locating the $S$-matrix poles is not an easy task, and especially for
few-body systems. There are different approaches to this
problem. To the best of our knowledge, so far only one of them has been
applied to study the hyperon-nucleus resonant states. This was done in
Ref. \cite{afnan} using an analytic continuation of the rigorous
three-body equations proposed by Alt, Grassberger, and Sandhas
\cite{AGS} and known as the AGS-equations. In our present paper, we
follow a different approach based on direct calculation of the Jost
function using the method suggested in Ref. \cite{nuovocim}.

The three-body systems we consider in the present paper, namely,
$\Lambda nn$ and $\Lambda\Lambda n$, do not have bound states in any
of the two-body subsystems $nn$, $\Lambda n$, or $\Lambda\Lambda$.
The only possible collision process for them is therefore the $3\to3$
scattering.  The wave functions describing the systems that cannot
form clusters behave asymptotically as linear combinations of the
incoming and outgoing hyperspherical waves (see, for example,
Ref. \cite{shmidt}). Thus it is convenient to describe the three-body
configuration using the hyperspherical coordinates, among which only
one (the hyperradius) runs from zero to infinity while all the others
(the hyperangles) vary within finite ranges.

Within the hyperspherical approach, the wave function is expanded in
an infinite series over the hyperspherical harmonics (similarly to the
partial wave decomposition in the two-body problem), and we end up
with an infinite system of coupled hyperradial equations, which is
truncated in practical calculations. All the details of the
hyperspherical approach can be found, for example, in the review by
M. Fabre de la Ripelle \cite{fabre}.

It should be noted that although the two-body potentials and masses
for the three-body systems $\Lambda nn$ and $\Lambda\Lambda n$ are
different, they can be treated using exactly the same
equations. Indeed, in both of these systems, we have two identical
particles with spin 1/2 and a third particle of the same spin. In what
follows, we therefore consider a general system of this type.

Let $m_1$ be the mass of one of the identical particles, and $m_2$ be
the mass of the third particle. Then the total mass of the system is
$M=2m_1+m_2$ and the reduced masses for the identical pair and for the
third particle are $\mu_1=m_1/2$ and $\mu_2=2m_1m_2/M$, respectively.
With the Jacobi coordinates shown in Fig. \ref{fig.jacobi},
the three-body Schr\"odinger equation can be written as
\begin{equation}
\label{Rak.HSeq}
    \left(
    \partial^2_r+\frac{5}{r}\partial_r-\frac{1}{r^2}{\cal L}^2+k^2-V
    \right)\Psi^{[s]}_{\vec k_1, \vec k_2}(\vec r_1, \vec r_2)=0\ ,
\end{equation}
where
\begin{equation}
\label{Rak.Vsum}
    V=2M(U_{12}+U_{13}+U_{23})
\end{equation}
is the sum of the two-body potentials $U_{ij}$, the vectors $\{\vec
k_1, \vec k_2\}$ represent the incident momenta of the three-body collision
along the corresponding configuration vectors $\{\vec r_1, \vec
r_2\}$, the superscript $[s]=((s_1s_2)s_{12}s_3)s\sigma$ denotes the
spin quantum numbers for the spin-addition scheme
$\vec{s}=(\vec{s_1}+\vec{s_2})+\vec{s_3}$, the variable
\begin{equation}
\label{Rak.hyperradius}
     r=\sqrt{r_1^2+r_2^2}\ ,
\end{equation}
is the hyperradius that gives the ``collective'' size of the system,
$k$ is related to the total energy, $k^2=2ME$, and can be called the
hypermomentum, and the operator ${\cal L}^2$ absorbs all the angular
variables. It is defined as
\begin{equation}
\label{Rak.hyperL}
     {\cal L}^2=
     -\frac{\partial^2}{\partial\alpha^2}-
     4\cot(2\alpha)\frac{\partial}{\partial\alpha}+
     \frac{1}{\cos^2\alpha}\vec{\ell}^2_{\vec r_1}+
     \frac{1}{\sin^2\alpha}\vec{\ell}^2_{\vec r_2}
\end{equation}
with $\alpha=\arctan(r_2/r_1),\quad 0\le\alpha\le\pi/2$, and
$\vec{\ell}_{\vec r_i}$ being the operators of the angular momenta
associated with the corresponding Jacobi coordinates. The solutions of
the eigenvalue problem
\begin{equation}
\label{Rak.hyperLeigen}
      {\cal L}^2Y_{[L]}(\omega)=L(L+4)Y_{[L]}(\omega)
\end{equation}
are the so called hyperspherical harmonics that depend on the
hyperangles $\omega=\{\Omega_{\vec r_1},\Omega_{\vec r_2},\alpha\}$
including the spherical angles $\Omega_{\vec r_i}$ of the vectors
$\vec r_i$ and the angle $\alpha$ that determines the ratio
$r_2/r_1$.  The subscript $[L]$ is the multi-index
$[L]=\{L,\ell_1,\ell_2,\ell,m\}$ that includes the grand orbital
quantum number,
\begin{equation}
\label{Rak.grand}
      L=\ell_1+\ell_2+2n\ ,\qquad n=0,1,2,\dots\ ,
\end{equation}
as well as the angular momenta associated with the Jacobi vectors and
the total angular momentum $\ell$ together with its third component
$m$.  Combining $Y_{[L]}(\omega)$ with the spin states
$\chi_{[s]}=|((s_1s_2)s_{12}s_3)s\sigma\rangle$, we obtain the
functions
\begin{equation}
\label{Rak.HHjj}
      \Phi_{[L]}^{jj_z}(\omega)=
      \sum_{m\sigma} \langle \ell m s \sigma | jj_z\rangle
      Y_{[L]}(\omega)\chi_{[s]}
\end{equation}
that constitute a full ortho-normal set of states with a given total angular
momentum $j$ in the spin-angular subspace.

Similarly to the two-body partial wave decomposition, we can expand
a solution of Eq. (\ref{Rak.HSeq}) in the infinite series over the
hyperspherical harmonics,
\begin{widetext}
\begin{equation}
\label{Rak.psiexpansion}
   \Psi_{\vec k_1, \vec k_2}^{[s]}(\vec r_1,\vec r_2)=
   \displaystyle\frac{1}{r^{5/2}}
   \sum_{[L][L']jj_z} u_{[L][L']}^{jj_z}(E,r)
   \Phi^{jj_z}_{[L]}(\omega_{\vec r})
   \Phi^{jj_z*}_{[L']}(\omega_{\vec k})\ ,
\end{equation}
\end{widetext}
where the hyperangle sets $\omega_{\vec r}$ and $\omega_{\vec k}$ are
associated with the pairs $\{\vec{r_1},\vec{r_2}\}$ and
$\{\vec{k_1},\vec{k_2}\}$, respectively.  After substituting this
expansion into Eq. (\ref{Rak.HSeq}) and doing the projection onto the
functions $\Phi^{jj_z}_{[L]}$, we end up with the following system of
hyperradial equations
\begin{equation}
\label{Rak.HRsystem}
    \left[\partial^2_r+k^2-\displaystyle
    \frac{\lambda(\lambda+1)}{r^2}\right]
    u_{[L][L']}=\displaystyle
    \sum_{[L'']}V_{[L][L'']}u_{[L''][L']}\ ,
\end{equation}
where for the sake of simplicity we dropped the superscripts $jj_z$
(indicating the conserving total angular momentum). In
Eq. (\ref{Rak.HRsystem}),
\begin{equation}
\label{Rak.Vmatr}
    V_{[L][L']}(r)=2M\int \Phi^{jj_z*}_{[L]}(\omega)
    \left(U_{12}+U_{13}+U_{23}\right)
    \Phi^{jj_z}_{[L']}(\omega)d\omega\ ,
\end{equation}
and $\lambda=L+3/2$.
Since we consider a system that cannot form clusters, the
asymptotic behaviour of its wave function may only involve the
incoming and outgoing hyperspherical waves $\sim \exp(\mp ikr)$,
which are the products of the corresponding spherical waves along the
Jacobi radii $r_1$ and $r_2$,
$$
    e^{ik_1r_1}e^{ik_2r_2}=e^{ikr\cos^2\alpha}e^{ikr\sin^2\alpha}
    =e^{ikr}\ .
$$
We therefore look for the solution of matrix equation (\ref{Rak.HRsystem})
as
\begin{widetext}
\begin{equation}
\label{Rak.HRansatz}
    u_{[L][L']}(E,r)=h_\lambda^{(-)}(kr)F_{[L][L']}^{(\rm in)}(E,r)
    +h_\lambda^{(+)}(kr)F_{[L][L']}^{(\rm out)}(E,r)\ ,
\end{equation}
\end{widetext}
where the incoming and outgoing hyperspherical waves described by the
Riccati-Hankel functions,
\begin{equation}
\label{Rak.hass}
    h_\lambda^{(\pm)}(kr)\,\,\mathop{\longrightarrow}_{|kr|\to\infty}\,\,
    \mp i\exp\left[\pm i(kr-\lambda\pi/2)\right]\ ,
\end{equation}
are included explicitly. The matrices $F_{[L][L']}^{(\rm
in/out)}(E,r)$ are new unknown functions. In the theory of
ordinary differential equations, this way of finding solution is known
as the variation parameters method (see, for example, Ref. \cite{brand}).

Since instead of one unknown matrix $u_{[L][L']}$ we introduce two
unknown matrices $F_{[L][L']}^{(\rm in/out)}$, they cannot be
independent. We therefore can impose an arbitrary condition that
relates them to each other. As such condition, it is convenient to
choose the following equation
\begin{equation}
\label{Rak.HRlagrange}
    h_\lambda^{(-)}(kr)\partial_rF_{[L][L']}^{(\rm in)}(E,r)
    +h_\lambda^{(+)}(kr)\partial_rF_{[L][L']}^{(\rm out)}(E,r)
    =0\ ,
\end{equation}
which is standard in the variation parameters method and is called the
Lagrange condition.  Substituting the ansatz (\ref{Rak.HRansatz}) into
the hyperradial equation (\ref{Rak.HRsystem}) and using the condition
(\ref{Rak.HRlagrange}), we obtain the following system of first order
equations for these unknown matrices
\begin{widetext}
\begin{equation}
\label{Rak.HRfinouteq}
    \left\{
    \begin{array}{rcl}
    \partial_rF_{[L][L']}^{(\rm in)} &=&
    -\displaystyle\frac{h_\lambda^{(+)}}{2ik}\sum\limits_{[L'']}
    V_{[L][L'']}\left[h_{\lambda''}^{(-)}F_{[L''][L']}^{(\rm in)}
    +h_{\lambda''}^{(+)}F_{[L''][L']}^{(\rm out)}\right]\ ,\\[3mm]
    \partial_rF_{[L][L']}^{(\rm out)} &=&
    +\displaystyle\frac{h_\lambda^{(-)}}{2ik}\sum\limits_{[L'']}
    V_{[L][L'']}\left[h_{\lambda''}^{(-)}F_{[L''][L']}^{(\rm in)}
    +h_{\lambda''}^{(+)}F_{[L''][L']}^{(\rm out)}\right]\ ,
    \end{array}
    \right.
\end{equation}
\end{widetext}
which are equivalent to the second order Eq. (\ref{Rak.HRsystem}). The
regularity of a physical wave function at $r=0$ implies the following
boundary conditions
\begin{equation}
\label{Rak.HRbcond}
    F_{[L][L']}^{(\rm in)}(E,0)= F_{[L][L']}^{(\rm out)}(E,0)
    =\delta_{[L][L']}\ .
\end{equation}
With these conditions, the columns of the matrix $u_{[L][L']}(E,r)$ are
not only regular but linearly independent as well. Therefore any
regular column $\phi_{[L]}(E,r)$ obeying Eq. (\ref{Rak.HRsystem}), can
be written as a linear combination of the columns of matrix
$u_{[L][L']}(E,r)$. In other words, the matrix $u_{[L][L']}(E,r)$ is a
complete basis for the regular solutions.

At large hyperradius where the potentials vanish, i.e.
\begin{equation}
\label{Rak.Vanish}
       V_{[L][L']}(r)
       \,\,\mathop{\longrightarrow}_{r\to\infty}\,\,0\ ,
\end{equation}
the right-hand sides of Eqs. (\ref{Rak.HRfinouteq}) should tend to
zero and therefore the matrices $F_{[L][L']}^{(\rm in/out)}(E,r)$
converge to the energy-dependent constants,
\begin{equation}
\label{Rak.HRjost}
    f_{[L][L']}^{(\rm in/out)}(E)=
    \lim\limits_{r\to\infty}F_{[L][L']}^{(\rm in/out)}(E,r),
\end{equation}
that by analogy with the two-body case can be called the Jost
matrices.  The convergency of these limits, however, depends on the
choice of the energy $E$ and on how fast the potential matrix
$V_{[L][L']}(r)$ vanishes when $r\to\infty$.

When the energy is real and positive (scattering states), the
vanishing of the right hand sides of Eqs. (\ref{Rak.HRfinouteq}) at
large distances is completely determined by the behaviour of
$V_{[L][L']}(r)$. It can be shown that in such a case the limits
(\ref{Rak.HRjost}) exist if $V_{[L][L']}(r)$ vanishes faster than
$1/r$.

With negative and complex energies there is a technical complication.
The problem is that one of the Riccati-Hankel functions on the right
hand side of Eqs. (\ref{Rak.HRfinouteq}) is always exponentially
diverging.  Therefore, if at large distances the potential matrix
vanishes not fast enough, the convergency of (\ref{Rak.HRjost}) is not
achieved.  This problem can be easily circumvented by using different
path to the far-away point (see Fig. \ref{Rak.fig.pathray}).

This is known as the the complex rotation of the coordinate.  All the
details concerning convergency of the limits (\ref{Rak.HRjost}) and
the use of complex rotation for this purpose can be found in
Refs. \cite{nuovocim,exactmethod,nnnn,partialwaves,singular,cplch}.

As was said before, the columns of the matrix function $u_{[L][L']}(E,r)$
constitute the regular basis using which we can construct a physical
solution $\phi_{[L]}(E,r)$ with given boundary conditions at infinity,
\begin{eqnarray}
\label{Rak.HRphys}
    \phi_{[L]}(E,r)=\sum_{[L']}u_{[L][L']}(E,r)C_{[L']}\ ,
\end{eqnarray}
where $C_{[L]}$ are the combination coefficients.

The spectral points $E_{\rm n}$ (bound and resonant states) are those
at which the physical solution has only outgoing waves in its
asymptotics, i.e. when
\begin{eqnarray}
\label{Rak.HRhomoeq}
    \sum_{[L']}f_{[L][L']}^{(\rm in)}(E_{\rm n})C_{[L']}=0\ .
\end{eqnarray}
This homogeneous system has a non-trivial solution if and only if
\begin{equation}
\label{Rak.HRdet0}
    \det f_{[L][L']}^{(\rm in)}(E_{\rm n})=0\ ,
\end{equation}
which determines the spectral energies $E_{\rm n}$. As can be easily
shown \cite{ijqc}, the $S$-matrix is given by
\begin{equation}
\label{Rak.Smatr}
   S(E)=f_\ell^{(\rm out)}(E)\left[f_\ell^{(\rm in)}(E)\right]^{-1}
\end{equation}
and therefore at the energies $E_{\rm n}$ it has poles.

\section{Two-body potentials}

In our calculations, we used local two-body
potentials describing the interaction between two neutrons, $\Lambda$
and neutron, and between two $\Lambda$-particles. For all these potentials,
we used the same functional form, namely,
\begin{widetext}
\begin{eqnarray}
\label{Rak.potform}
    U(\rho) &=& \left[A_1(\rho)-\frac{1+P^\sigma}{2}A_2(\rho)
    -\frac{1-P^\sigma}{2}A_3(\rho)\right]
    \left[\frac{\beta}{2}+\frac12(2-\beta)P^r\right]\ ,\\[3mm]
\label{Rak.potformA}
    A_n(\rho) &=& W_n\exp(-a_n\rho^2)\ ,\qquad n=1,2,3\ ,
\end{eqnarray}
\end{widetext}
where $P^\sigma$ and $P^r$ are the permutation operators in the spin
and configuration spaces, respectively. The form of $U(\rho)$ as well as
the parameters were taken from Ref. \cite{nemura}.  In order to
explore how sensitive the positions of the three-body resonances are
to the choice of underlying two-body potentials, we did the
calculations with three different sets of parameters for the $\Lambda
n$-potential. All the sets of parameters we used, are given in Table
\ref{Rak.table.parameters}.

\section{The minimal approximation}

The system (\ref{Rak.HRfinouteq}) consists of infinite number of
equations. For any practical calculation, one has to truncate it
somewhere. Before going any further, it is very logical to try the
simplest approximation, namely, when only the first terms of the sums
on the right hand sides of Eqs. (\ref{Rak.HRfinouteq}) are retained.
This corresponds to the minimal ($n=0$) value of the grand orbital
number (\ref{Rak.grand}) and is called the hypercentral approximation,
$[L]=[L_{\rm min}]$. We assume that the two-body subsystems are in the
$S$-wave states ($\ell_1=\ell_2=0$), which means that
$$
    \lambda=\lambda_{\rm min}=\displaystyle\frac32\ .
$$
So, in the minimal approximation, instead of the infinite system
(\ref{Rak.HRfinouteq}), we remain with only one equation,
\begin{equation}
\label{Rak.HRmin}
    \left[\partial^2_r+k^2-\displaystyle
    \frac{\lambda_{\rm min}(\lambda_{\rm min}+1)}{r^2}\right]
    u(E,r)=2M\langle U\rangle u(E,r)\ ,
\end{equation}
where all unnecessary subscripts are dropped, and the brackets on the
right hand side mean the following integration
\begin{equation}
\label{Rak.HRUav}
    \langle U\rangle(r)=
    \int \Phi^{jj_z*}_{[L_{\rm min}]}(\omega)
    \left(U_{12}+U_{13}+U_{23}\right)
    \Phi^{jj_z}_{[L_{\rm min}]}(\omega)d\omega\ .
\end{equation}
From the mathematical point of view, Eq. (\ref{Rak.HRmin}) looks exactly
like the two-body radial Schr\"odinger equation. The only difference
is that the angular momentum is not an integer number.

The explicit expression for the integral (\ref{Rak.HRUav}) is given in
the Appendix.  The hypercentral potentials $\langle U\rangle$ for the
systems $\Lambda nn$ and $\Lambda\Lambda n$ are shown in Figs
\ref{Rak.fig.LnnU} and \ref{Rak.fig.LLnU}.  With these hyperradial
potentials the corresponding differential equations determining the
three-body Jost functions, were numerically solved with complex values
of the energy. The results of these calculations are discussed next.

\section{Numerical results and conclusion}

When looking for zeros of the three-body Jost functions, we found
that there were no such zeros at real negative energies. In other words,
neither the system $\Lambda nn$ nor $\Lambda\Lambda n$ have bound
states.

The only zeros we found were located on the unphysical sheet of the
energy surface, in the resonance domain. The resonance energies are
given in Tables \ref{Rak.table.resonancesLnn} and
\ref{Rak.table.resonancesLLn} and shown in
Fig. \ref{Rak.fig.poles}. As is seen, the positions of these resonances
depend on the choice of the $\Lambda n$ potential. For the choice
``C'', the resonances become sub-threshold.

In order to estimate how far our three-body systems are from being
bound, we artificially increased the depths of the potentials by
multiplying them by a scaling factor. When this factor was increased
from 1 upwards, the Jost function zeros moved towards the origin of
the energy surface. At the value of approximately 1.5, the zeros
crossed the threshold and moved onto the real negative axis. In other
words, the bound states can appear if the potential strength is
increased by $\sim50$\%.

The fact that we did not find bound $\Lambda nn$ or $\Lambda\Lambda n$
states is not surprising at all. As is shown in Refs. \cite{miyagawa,
garcilazo}, the system $\Lambda NN$ in the state with the three-body
isospin 1 and spin $s=1/2$ is not bound even when the virtual
processes of $\Lambda-\Sigma$ conversion are taken into account,
although this conversion increases the attraction in the system.
Simple but convincing argumentation of Ref. \cite{tang} leads us to
the conclusion that the $\Lambda\Lambda n$ system also cannot be
bound.  Indeed, the system $\Lambda\Lambda n$ is a ``mirror'' image of
$\Lambda nn$, where the $\Lambda$ and $n$ replace each other. This
means that the potential term $U=U_{nn}+U_{\Lambda n}+U_{\Lambda n}$
of the three-body Hamiltonian is replaced with
$U=U_{\Lambda\Lambda}+U_{\Lambda n}+U_{\Lambda n}$.  Since the
attraction of $U_{\Lambda\Lambda}$ is weaker than that of $U_{nn}$, we
may conclude that the system $\Lambda\Lambda n$ has less chances to be
bound than the system $\Lambda nn$. The calculations performed in
Refs.  \cite{nemura, filikhin, nemuraLL, galLL}, show that even the heavier
hypernucleus ${}^{\phantom{\Lambda}4}_{\Lambda\Lambda}$H (i.e. the
system $\Lambda\Lambda pn$) is bound very weakly, if bound at all.

Multiplying the two-body potentials by an appropriate scaling factor,
we can always generate an artificial three-body bound state, i.e. a
pole of the $S$-matrix on the physical sheet of the $E$-surface at a
negative energy. Apparently this pole cannot disappear when the
scaling factor returns to its natural value of 1. The pole simply
moves via the threshold onto the unphysical sheet. Since both the
systems we consider, are not far from being bound, their corresponding
poles cannot be far away from the threshold energy. And indeed we
located them at low energies.

What we found is, of course, an estimate. But it clearly shows that
there are near-threshold resonances of the systems $\Lambda nn$ and
$\Lambda\Lambda n$.  Actual location of the poles most probably is
more close to the threshold energy. An inclusion of the channels
$\Lambda N-\Sigma N$ and $\Lambda\Lambda-\Xi N$ would definitely
increase the attraction in our systems (see Ref. \cite{hiyama}) and
this would make the widths of the resonances smaller.

As we have demonstrated, the positions of the resonances strongly
depend on the choice of the two-body potentials. If such resonances
are observed experimentally, they may serve as an additional
instrument for constructing adequate $YN$ and $YY$ potentials.  There
are many possible reactions where the three-body resonances $\Lambda
nn$ and $\Lambda\Lambda n$ may manifest themselves. As an example, we
can mention the inelastic collision of the $K^-$ meson with the
$\alpha$ particle,
\begin{equation}
\label{reaction}
   K^-+{}^4{\rm He}\ \longrightarrow\ p+\Lambda+n+n\ ,
\end{equation}
that produces a proton and the system we are looking for. If a
short-lived cluster $\Lambda nn$ is formed in the final state of this
collision, it should be seen in the corresponding two-body kinematics
$p-\Lambda nn$. The processes of the type (\ref{reaction}) fall under
the experimental programme AMADEUS \cite{amadeus} (in the INFN, Italy)
and, in principle, this reaction could be thoroughly studied.

\vspace{1cm}
\begin{appendix}
\begin{center}
{\bf APPENDIX}
\end{center}
Hypercentral potential (\ref{Rak.HRUav}) consists of the three terms
\begin{equation}
\label{Rak.HCpot}
    \langle U\rangle=\langle U_{12}\rangle+
                     \langle U_{13}\rangle+
                     \langle U_{23}\rangle\ ,
\end{equation}
where $U_{ij}$ is the two-body potential acting between particles $i$
and $j$.  As was mentioned above, we can consider both $nn\Lambda$ and
$\Lambda\Lambda n$ systems in a unified way.  Let 1 and 2 be the
identical particles, i.e. the $nn$ or $\Lambda\Lambda$ pair, and 3 be
the remaining $\Lambda$-particle or neutron, respectively.

The six-dimensional volume element is
\begin{widetext}
\begin{eqnarray*}
       d\vec{r}_1d\vec{r}_2 &=&
       r_1^2r_2^2dr_1dr_2
       \sin\theta_{\vec r_1}d\theta_{\vec r_1}d\varphi_{\vec r_1}
       \sin\theta_{\vec r_2}d\theta_{\vec r_2}d\varphi_{\vec
       r_2}\\[3mm]
       &=&
       r^5dr\frac14\sin^2(2\alpha)d\alpha
       \sin\theta_{\vec r_1}d\theta_{\vec r_1}d\varphi_{\vec r_1}
       \sin\theta_{\vec r_2}d\theta_{\vec r_2}d\varphi_{\vec
       r_2}\\[3mm]
       &=&
       r^5drd\omega\ .
\end{eqnarray*}
\end{widetext}
Therefore in the five-dimensional integral (\ref{Rak.HRUav}) the
volume element is
\begin{equation}
\label{Rak.domega}
      d\omega=\frac14\sin^2(2\alpha)
      \sin\theta_{\vec r_1}\sin\theta_{\vec r_2}
      d\alpha d\theta_{\vec r_1}d\varphi_{\vec r_1}
      d\theta_{\vec r_2}d\varphi_{\vec r_2}\ .
\end{equation}

Since we assume that $\ell_1=\ell_2=0$ and $L=L_{\rm min}=0$, the sum
(\ref{Rak.HHjj}) is reduced to a single term,
\begin{equation}
\label{Rak.HHjjmin}
      \Phi_{[L_{\rm min}]}^{jj_z}(\omega)=
      Y_{[L_{\rm min}]}(\omega)\chi_{[s]}\ ,
\end{equation}
where the quantum numbers $jj_z$ coincide with $s\sigma$. The two-body
spin $s_{12}$ of the identical pair in the $S$-wave state must be
zero. As a result the three-body spin $s$ is always 1/2.
The hyperspherical harmonics $Y_{[L_{\rm min}]}(\omega)$ is trivial
(independent of the angles),
\begin{equation}
\label{Rak.YLmin}
      Y_{[L_{\rm min}]}(\omega)\equiv \pi^{-3/2}\ ,
\end{equation}
which means that the action of the permutation operators $P^r$ for all
three terms in Eq. (\ref{Rak.HCpot}) is also trivial: its eigenvalue
is 1,
\begin{equation}
\label{Rak.PrYLmin}
      P^r_{ij}Y_{[L_{\rm min}]}=Y_{[L_{\rm min}]}\ ,\qquad
      ij=\{12\},\{13\},\{23\}\ .
\end{equation}
The spin permutation operator $P^\sigma_{12}$ for the identical pair
\{12\} changes the sign of $\chi_{[s]}$,
\begin{equation}
\label{Rak.Ps12Chi}
      P^\sigma_{12}\chi_{[s]}=-\chi_{[s]}\ ,
\end{equation}
because $s_{12}=0$ in $[s]=((s_1s_2)s_{12}s_3)s\sigma$. For the other
two pairs, its action is a bit more complicated. Indeed, recoupling
the spins,
\begin{widetext}
\begin{eqnarray*}
     |((s_1s_2)s_{12}s_3)s\sigma\rangle =
     \sum_{s_{31}}
     |((s_3s_1)s_{31}s_2)s\sigma\rangle
     \langle ((s_3s_1)s_{31}s_2)s\sigma|
     ((s_1s_2)s_{12}s_3)s\sigma\rangle\\[3mm]
     =
     |((s_3s_1)0s_2)s\sigma\rangle
     \left\{
     \begin{array}{ccc}
     \frac12 & \frac12 & 0\\
     \frac12 & \frac12 & 0
     \end{array}
     \right\}+
     \sqrt{3}|((s_3s_1)1s_2)s\sigma\rangle
     \left\{
     \begin{array}{ccc}
     \frac12 & \frac12 & 0\\
     \frac12 & \frac12 & 1
     \end{array}
     \right\}\\[3mm]
     =
     -\frac{1}{2}|((s_3s_1)0s_2)s\sigma\rangle+
     \frac{\sqrt{3}}{2}|((s_3s_1)1s_2)s\sigma\rangle\ ,
\end{eqnarray*}
\end{widetext}
we find that
\begin{equation}
\label{Rak.Ps13Chi}
      \chi_{[s]}^+P^\sigma_{13}\chi_{[s]}=\frac12\ .
\end{equation}
Similarly, it is easy to find for the pair \{23\} that
\begin{equation}
\label{Rak.Ps23Chi}
      \chi_{[s]}^+P^\sigma_{23}\chi_{[s]}=\frac12\ .
\end{equation}
When inserting the potentials $U_{ij}$ given by
Eq. (\ref{Rak.potform}), into the integral (\ref{Rak.HRUav}),
we should use the following interparticle distances (see
Fig. \ref{fig.jacobi}),
\begin{widetext}
\begin{eqnarray}
\label{Rak.rho12}
       \rho_{12} &=& r\sqrt{\frac{M}{\mu_1}}\cos\alpha\ ,\\[3mm]
\label{Rak.rho13}
       \rho_{13} &=& r\sqrt{\frac{M}{\mu_2}\sin^2\alpha+
                     \frac{M}{4\mu_1}\cos^2\alpha-
                     \frac{M}{2\sqrt{\mu_1\mu_2}}
                     \sin(2\alpha)\cos\theta_{\vec r_2}}\ ,\\[3mm]
\label{Rak.rho23}
       \rho_{23} &=& r\sqrt{\frac{M}{\mu_2}\sin^2\alpha+
                     \frac{M}{4\mu_1}\cos^2\alpha+
                     \frac{M}{2\sqrt{\mu_1\mu_2}}
                     \sin(2\alpha)\cos\theta_{\vec r_2}}\ .
\end{eqnarray}
\end{widetext}
Since the particles 1 and 2 are identical the interactions $U_{13}$
and $U_{23}$ are the same. Moreover, according to
Eqs. (\ref{Rak.Ps13Chi}, \ref{Rak.Ps23Chi}), the product
$\chi_{[s]}^+U_{13}\chi_{[s]}$ has the same dependence on $\rho_{13}$
as $\chi_{[s]}^+U_{23}\chi_{[s]}$ depends on $\rho_{23}$. Actually,
they depend on $\rho^2_{13}$ and $\rho^2_{23}$ because all the terms
in the potential (\ref{Rak.potform}) are of the Gaussian form.
Comparing Eqs. (\ref{Rak.rho13}) and (\ref{Rak.rho23}), we see that
the integrands for $\langle U_{13}\rangle$ and $\langle U_{23}\rangle$
differ only in the sign of the power of the exponential factors
corresponding to the last terms of $\rho^2_{13}$ and $\rho^2_{23}$.
This difference however has no effect on the integrals. Indeed,
the integration over $\theta_{\vec r_2}$,
\begin{widetext}
\begin{eqnarray*}
     \int_0^\pi\exp(\pm f\cos\theta_{\vec r_2})
     \sin\theta_{\vec r_2}d\theta_{\vec r_2}
     &=&
     \int_{-1}^{1}\exp(\pm ft)dt\\[3mm]
     &=&
     \frac{1}{f}(e^f-e^{-f})
     =
     \frac{2}{f}\sinh(f)\ ,
\end{eqnarray*}
gives the same result for both signs. Therefore
$\langle U_{13}\rangle=\langle U_{23}\rangle$ and hence
\begin{equation}
\label{Rak.HCpot1}
    \langle U\rangle=\langle U_{12}\rangle+
                     2\langle U_{13}\rangle\ .
\end{equation}
Performing trivial integrations over $\varphi_{\vec r_1}$,
$\varphi_{\vec r_2}$, $\theta_{\vec r_1}$, and $\theta_{\vec r_2}$
(trivial in the case of $\langle U_{12}\rangle$), we obtain the
following expressions for the terms of the hypercentral potential
(\ref{Rak.HCpot1}),
\begin{eqnarray}
\label{Rak.W11}
    \langle U_{12}\rangle &=&
    \frac{4}{\pi}\int_0^{\pi/2}d\alpha
    \sin^2(2\alpha)\left[W_1^{\{12\}}\exp\left(-a_1^{\{12\}}
    \eta r^2\right)
    -W_3^{\{12\}}\exp\left(-a_3^{\{12\}}\eta r^2\right)
    \right]\ ,\\[3mm]
\label{Rak.W13}
    \langle U_{13}\rangle &=&
    \frac{2}{\pi}\int_0^{\pi/2}d\alpha
    \sin^2(2\alpha)\left[W_1^{\{13\}}\exp\left(-a_1^{\{13\}}\zeta
    r^2\right)s\left(a_1^{\{13\}}\xi r^2\right)\right.\\[3mm]
\nonumber
    &&\hspace{31mm}
    -\frac34W_2^{\{13\}}\exp\left(-a_2^{\{13\}}\zeta r^2\right)
    s\left(a_2^{\{13\}}\xi r^2\right)\\[3mm]
\nonumber
    &&\hspace{31mm}
    \left.
    -\frac14W_3^{\{13\}}\exp\left(-a_3^{\{13\}}\zeta r^2\right)
    s\left(a_3^{\{13\}}\xi r^2\right)
    \right]\ ,
\end{eqnarray}
\end{widetext}
where
\begin{eqnarray*}
    \eta(\alpha) &=& \frac{M}{\mu_1}\cos^2\alpha\ ,\\[3mm]
    \zeta(\alpha)&=& \frac{M}{\mu_2}\sin^2\alpha
                     +\frac{M}{4\mu_1}\cos^2\alpha\ ,\\[3mm]
    \xi(\alpha)  &=&
    \frac{M}{2\sqrt{\mu_1\mu_2}}\sin(2\alpha)\ ,\\[3mm]
    s(f)  &=& \frac{1}{f}\left(e^f-e^{-f}\right)\ .
\end{eqnarray*}
The parameters $W_n^{\{ij\}}$ and $a_n^{\{ij\}}$, where the symbol $\{ij\}$
means a choice of the pair of interacting particles, are given in
Table \ref{Rak.table.parameters}.  For each (complex) value of the
hyperradius $r$, which was needed in our calculations, the integrals
(\ref{Rak.W11}) and (\ref{Rak.W13}) were evaluated numerically.
\end{appendix}
\begin{acknowledgments}
Financial support from Deutsche Forschungsgemeinschaft (DFG grant no
436 RUS 113/761/0-2) is greatly appreciated.
\end{acknowledgments}


\newpage
\begin{widetext}
\begin{center}
\begin{table}
\begin{center}
\begin{tabular}{c|c|c|c|c|c}
 & $nn$ & $\Lambda\Lambda$ & $\Lambda n$\,(A) & $\Lambda n$\,(B)
 & $\Lambda n$\,(C)\\
\hline
$W_1$\,(MeV)       & 200.0 &  200.0 &  200.0  &  600.0  &  5000  \\[1mm]
$W_2$\,(MeV)       & 178.0 &  0     &  106.5  &  52.61  &  47.87 \\[1mm]
$W_3$\,(MeV)       & 91.85 &  130.8 &  118.65 &  66.22  &  61.66 \\[1mm]
$a_1$\,(fm$^{-2}$) & 1.487 &  2.776 &  1.638  &  5.824  &  18.04 \\[1mm]
$a_2$\,(fm$^{-2}$) & 0.639 &  0     &  0.7864 &  0.6582 &  0.6399\\[1mm]
$a_3$\,(fm$^{-2}$) & 0.465 &  1.062 &  0.7513 &  0.6460 &  0.6325\\[1mm]
$\beta$            & 1     &  1     &  1      &  1      &      1 \\[1mm]
\hline
\end{tabular}
\end{center}
\caption{Parameters of the potential (\ref{Rak.potform}) for the pairs
  $nn$, $\Lambda\Lambda$, and $\Lambda n$. For the system $\Lambda n$,
  three different sets of parameters (denoted as A, B, and C) are
  given. All the parameters are taken from Ref. \cite{nemura}.
}
\label{Rak.table.parameters}
\end{table}
\begin{table}
\begin{center}
\begin{tabular}{c|c|c|c}
$\Lambda n$-potential & A & B & C\\
\hline
$E_0$, (MeV) &
$0.551-\displaystyle\frac{i^{\mathstrut}}{2_{\mathstrut}}4.698$ &
$0.456-\displaystyle\frac{i}{2}4.885$ &
$-0.149-\displaystyle\frac{i}{2}5.783$
\end{tabular}
\end{center}
\caption{Complex resonance energies
$E_0=E_{\rm r}-\displaystyle\frac{i}{2}\Gamma$
for the system $\Lambda nn$ with
  the three choices of $\Lambda n$-potential.}
\label{Rak.table.resonancesLnn}
\end{table}
\begin{table}
\begin{center}
\begin{tabular}{c|c|c|c}
$\Lambda n$-potential & A & B & C\\
\hline
$E_0$, (MeV) &
$0.096-\displaystyle\frac{i^{\mathstrut}}{2_{\mathstrut}}8.392$ &
$0.034-\displaystyle\frac{i}{2}8.438$ &
$-0.552-\displaystyle\frac{i}{2}8.681$
\end{tabular}
\end{center}
\caption{Complex resonance energies
$E_0=E_{\rm r}-\displaystyle\frac{i}{2}\Gamma$
for the system $\Lambda\Lambda n$ with
  the three choices of $\Lambda n$-potential.}
\label{Rak.table.resonancesLLn}
\end{table}
\begin{figure}
\centerline{\epsfig{file=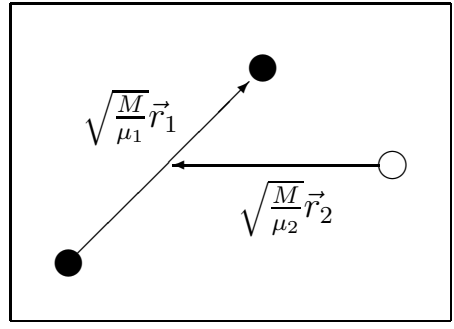,width=60mm}}
\caption{\sf
Jacobi vectors defining the spatial configuration of a three-body
system of two identical (filled circles) and one different (open
circle) particles.
}
\label{fig.jacobi}
\end{figure}
\begin{figure}
\centerline{\epsfig{file=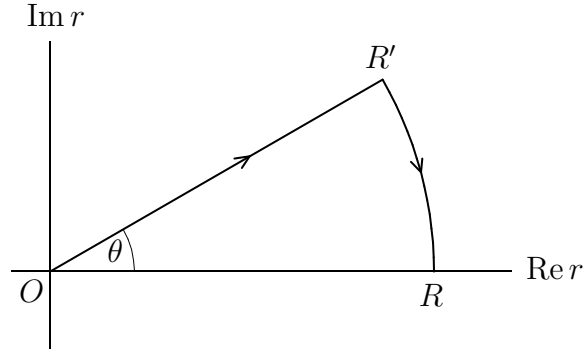,width=80mm}}
\caption{\sf
Deformed contour for integrating Eqs. (\ref{Rak.HRfinouteq}) from $r=0$
to $r=R$ when the energy is complex.
}
\label{Rak.fig.pathray}
\end{figure}
\begin{figure}
\centerline{\epsfig{file=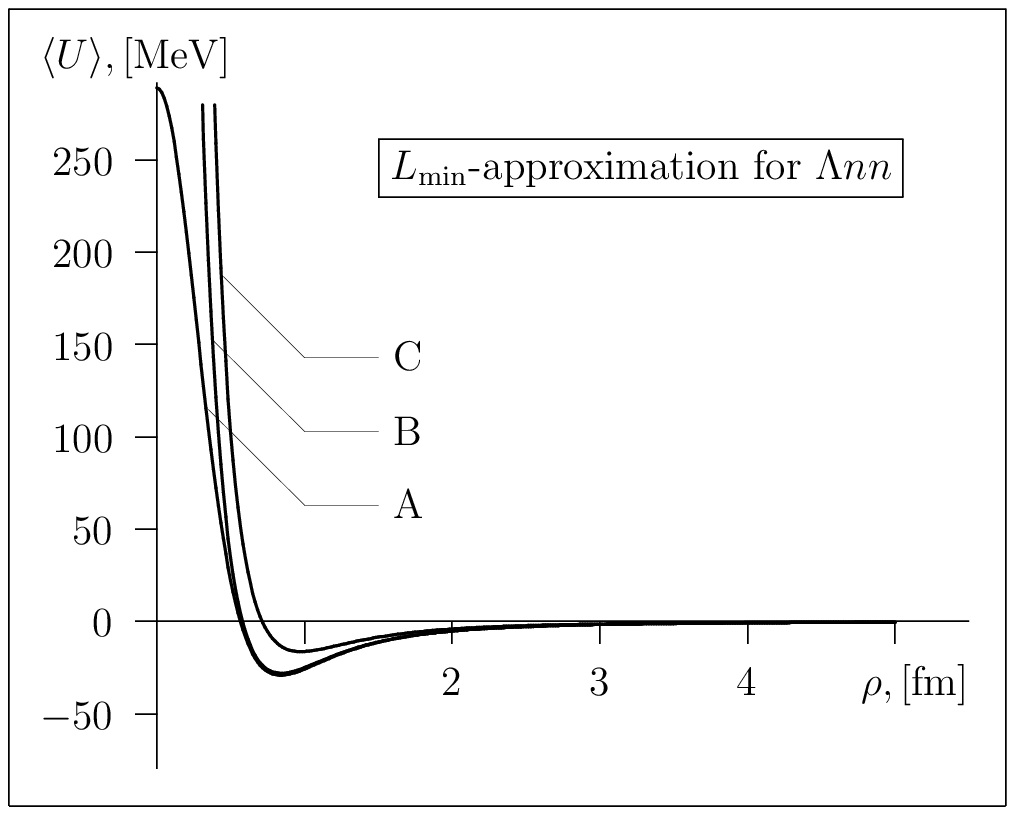,width=110mm}}
\caption{\sf
The hypercentral potential given by Eq. (\ref{Rak.HRUav}) for the
system $\Lambda nn$ with the three
choices (A, B, and C) of the $\Lambda n$ interaction.
}
\label{Rak.fig.LnnU}
\end{figure}
\begin{figure}
\centerline{\epsfig{file=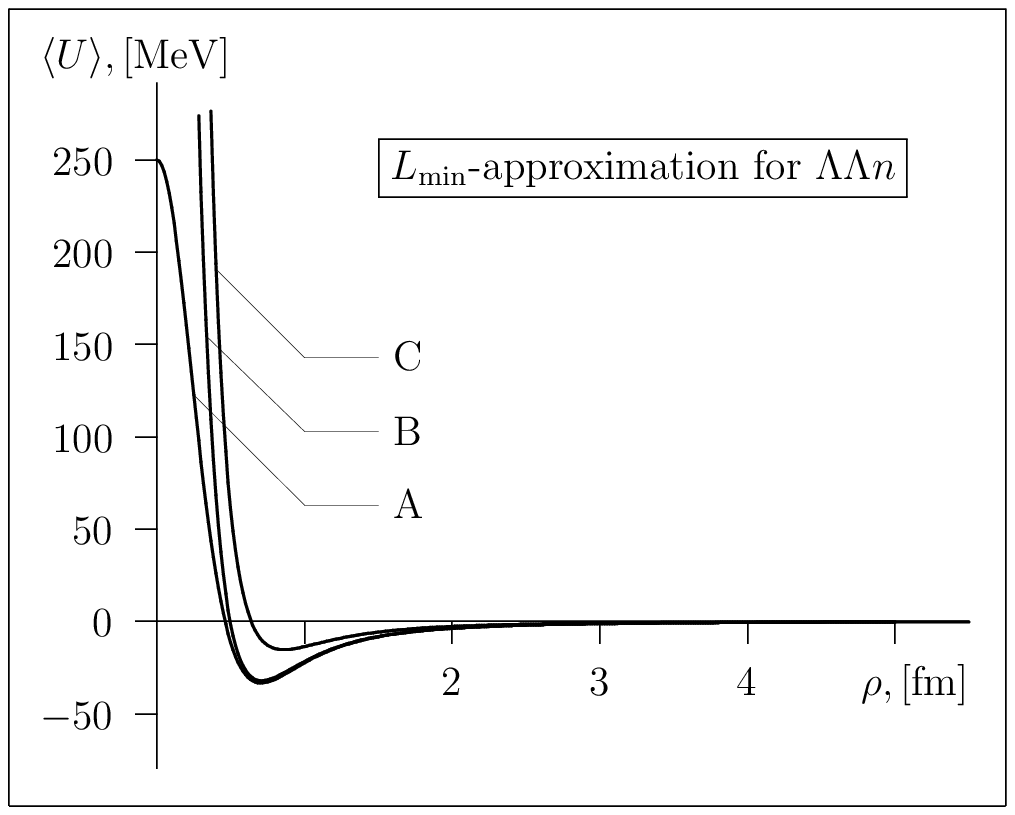,width=110mm}}
\caption{\sf
The hypercentral potential given by Eq. (\ref{Rak.HRUav}) for the
system $\Lambda\Lambda n$ with the three
choices (A, B, and C) of the $\Lambda n$ interaction.
}
\label{Rak.fig.LLnU}
\end{figure}
\begin{figure}
\centerline{\epsfig{file=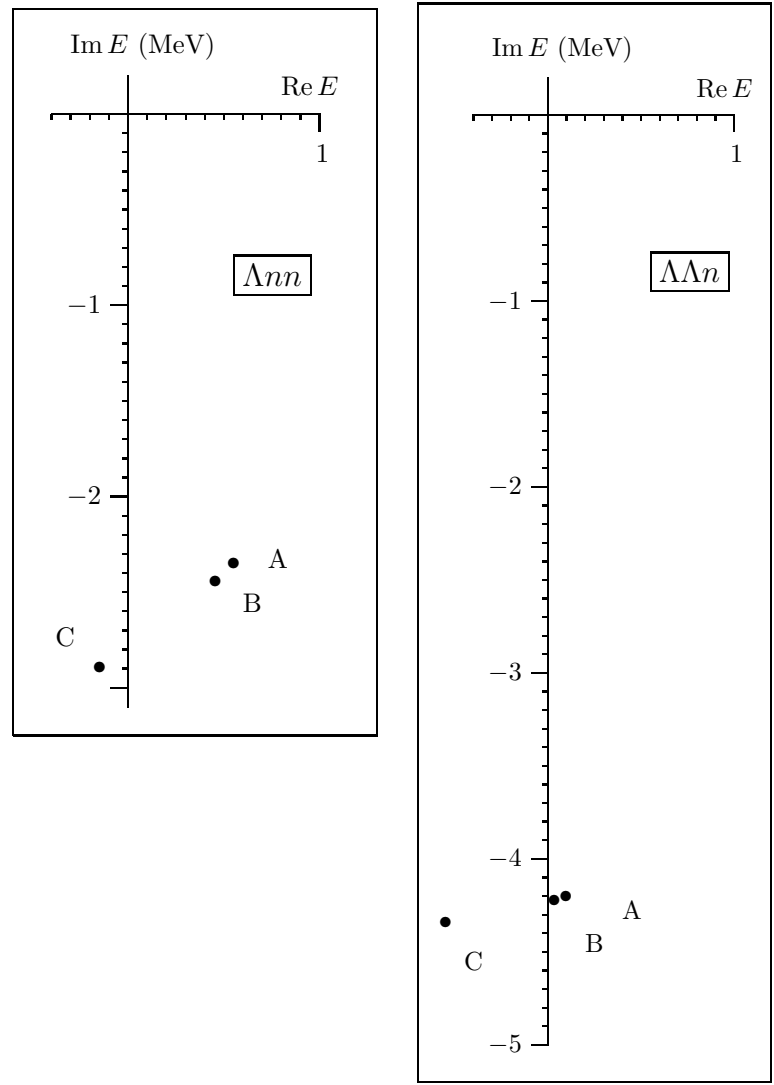,width=105mm}}
\caption{\sf
Resonance points for the systems $\Lambda nn$ and $\Lambda\Lambda n$
found on the unphysical sheet of the energy surface
with the three sets (A, B, and C) of parameters of the $\Lambda
n$-potential given in Table \ref{Rak.table.parameters}.
}
\label{Rak.fig.poles}
\end{figure}
\end{center}
\end{widetext}
\end{document}